\documentclass[10pt, draftcls, onecolumn]{IEEEtran}
\usepackage {graphicx,alltt,epsf}
\usepackage{times, amsmath, amssymb, epsfig}
\usepackage[hyphens]{url}
\usepackage[T1]{fontenc}
\usepackage{bm}
\usepackage{multirow}
\usepackage{subfigure}

\begin{document}
%\title{Using the Style File IEEEtran.sty}
\title{Improved Robust Node Position Estimation in Wireless Sensor Networks} %!PN
\author{R.~C.~Nongpiur
 % stops a space
%\vspace{-0.2in}      
\thanks{R.~C.~Nongpiur is with the Department
of Electrical and Computer Engineering, University of Victoria, Victoria,
BC, Canada V8W 3P6 e-mail: rnongpiu@ece.uvic.ca}% <-this % stops a space
}

\maketitle
\begin{abstract}
A new method for estimating the relative positions of location-unaware nodes from the location-aware nodes and the received signal strength (RSS) between the nodes, in a wireless sensor network (WSN), is proposed. In the method, a regularization term is incorporated in the optimization problem leading to significant improvement in the estimation accuracy even in the presence of position errors of the location-aware nodes and distance errors between the nodes. The regularization term is appropriated weighted on the basis of the degree of connectivity between the nodes in the network. The method is formulated as a convex optimization problem using the semidefinite relaxation approach. Experimental comparisons with state-of-the-art competing methods show that the proposed method yields node positions that are much more accurate even in the presence of measurement errors.
\end{abstract}

\begin{IEEEkeywords}
wireless sensor networks, robust node position estimation, received signal strength
\end{IEEEkeywords}
%\vspace{-0.1in}
\section{Introduction}
In WSNs~\cite{aky}, knowledge of the positions of the sensor nodes is required for most sensing tasks  such as enhancing the efficiency of routing protocol~\cite{ko}, localization and tracking~\cite{bahl}, and node subset selection~\cite{chen}, to name a few. Though a node can be made position aware by incorporating a global positioning system (GPS) unit or by presetting with location information, the two approaches have their own drawbacks. In the former, including a GPS unit in all the nodes would significantly increase the cost and power consumption~\cite{moses} of the WSN, while in the latter, the calibration of position information for each node would slow down the deployment process and would constrain the nodes to fixed positions. A more feasible approach is to have a limited number of location-aware (LA) nodes that would facilitate the location-unaware (LU) nodes to estimate their relative positions~\cite{savvides}.
 
In general, there are three popular measurement information that can be used to estimate the node positions, namely, time of arrival~\cite{guvenc, venkatesh}, time difference of arrival~\cite{catovic}-\cite{sun}, angle of arrival~\cite{vaghefi, cong}, and RSS~\cite{patwari1}-\cite{so}. Among the three, the RSS information is most popular due to simplicity and lower cost~\cite{patwari2}. In this paper, we consider localization based on RSS information.

The problem of localization of the sensor nodes can be classified as cooperative or non-cooperative~\cite{patwari2}. In non-cooperative localization, only measurements between the LA nodes and the LU nodes are used for position estimation, while in cooperative localization, the measurement between the LU nodes are also used. The additional information gained from the measurements between the LU nodes enhances the accuracy and robustness of the localization algorithm. Nowadays, most localization algorithms for WSNs are based on cooperative localization~\cite{yang}-\cite{vaghefi2}. 

Recent efforts to address the node localization problem have focused on optimization methods~\cite{yang, ho}. Since the work in~\cite{doherty}, several methods based on optimization have appeared in the literature. In~\cite{biswas}, a method based on semidefinite relaxation (SDR) followed by a gradient descent approach for refinement was proposed. Then in~\cite{tseng}, a method based on second-order cone programming (SOCP) relaxation was developed. Though the method works well as long as the LU nodes lie within the convex hull of the location, it performance  deteriorate as the number of LU nodes outside the convex hull increases~\cite{yang}. More recently, in~\cite{chiu} and~\cite{vaghefi2} the SDR approach was adopted to solve the node localization problem. While the method in~\cite{vaghefi2} considered a WSN having nodes with unknown transmit powers, the method in~\cite{chiu} considered a WNS where the positions of the LA nodes are inexact and the RSS information is subjected to fading.

The localization problem has also been approached using maximum likelihood (ML) estimation methods~\cite{patwari1, chang, moses}. A drawback of the ML estimation methods is that the cost function of the estimator is highly nonlinear and nonconvex and the quality of the final solution is very much dependent on the initial solution. To obtain good initial solutions, initial various approaches such as grid search, linear estimators, and convex relaxation have been used~\cite{vaghefi2}.

In this paper, we propose a new method for estimating the position of LU nodes using the positions of the LA nodes and the RSS information shared between the nodes. In the method, a regularization term is incorporated in the optimization problem leading to significant improvement in the estimation accuracy even in the presence of
position errors of the location-aware nodes and distance errors between the nodes. The regularization term is appropriately weighted on the basis of the degree of connectivity between the nodes in the network. The method is formulated as a convex optimization problem using the SDR approach. Experimental comparisons with state-of-the-art competing methods show that the proposed method yields node positions that are much more accurate.

The paper is organized as follows. In Section II, we describe the position estimation problem and associated formulations for imperfect node-positions and RSS with fading. Then in Section III, we develop formulations for solving the optimization problem. In Section IV, performance comparisons between the proposed method and state-of-the-art competing methods are carried out. Conclusions are drawn in Section V.

\section{Problem Statement}
We consider a WSN scenario where there are $N$ LU nodes and $M$ LA nodes. Let $\mathbf{x}_n$ and $\mathbf{a}_m$ be two-element vectors that correspond to the two-dimensional coordinates of the $n$th LU and $m$th LA nodes, respectively. We assume that the coordinates of the LA nodes have inaccuracies due to measurement errors~\cite{yang, chiu}. If $\mathbf{\bar{a}}_m$ is the inexact coordinate of the $m$th LA node, the relation between $\mathbf{a}_m$ and $\mathbf{\bar{a}}_m$ is given by
\begin{equation}
\mathbf{\bar{a}}_m = \mathbf{a}_m + {\bm \delta}_m
\end{equation}
where
\begin{equation}
\|{\bm \delta}_m\|_2 < \epsilon
\end{equation}
and $\epsilon$ is the $L_2$ norm bound of the error.

In conventional estimation problems, it is typical to compare the proposed method with the theoretical Cramer-Rao bound (CRB)~\cite{vanTrees} by assuming a certain distribution for the measurements, like for e.g., a normal distribution model. However, in this paper we assume that the measurement errors have unknown statistical properties and, therefore, the CRB cannot be derived. Furthermore, the optimization problem proposed in this paper cannot be proven to be unbiased and, consequently, even if the CRB is derived by assuming a certain distribution the results could be misleading. Due to these reasons, the CRB analysis is not included in this paper.

As in~\cite{patwari1, patwari2}, we assume that the RSS is subjected to fading. If $p_0$ is the received power at reference $d_0$, the estimated distance $\bar{d}_{ij}$ between sensor $i$ and sensor $j$ in the presence of large-scale fading can be modeled as~\cite{patwari1}
\begin{equation}
\bar{d}_{ij} = d_0 10^{\frac{p_0 - \bar{p}_{ij}}{10 \gamma_p}} = d_{ij} 10^{\frac{\alpha_{ij}}{10\gamma_p}}
\label{distFading}
\end{equation}
where
\begin{eqnarray}
\bar{p}_{ij} & = & p_{ij} + \alpha_{ij} \\
p_{ij} & = & p_0 - 10\gamma_p \log \left( \frac{d_{ij}}{d_0} \right)
\end{eqnarray}
$\gamma_p$ is the path loss exponent, $d_{ij}$ is the actual distance between the sensors, $\bar{p}_{ij}$ is the measured power in dB, $p_{ij}$ is the mean power in dB, and $\alpha_{ij}$ is the fading gain, which is normally distributed with zero mean and variance $\sigma_{dB}^2$.

As in~\cite{chiu, biswas}, we assume the realistic scenario where the distance measurements are affected by limitations in ranging, so that only inter-node distances less than $d_{max}$ can be measured. Consequently, for $n = 1, \ldots , N$, we define the set $\mathcal{N}(n)$ as
\begin{equation}
\mathcal{N}(n) = \mathcal{N}_1(n) \cup \mathcal{N}_2(n)
\end{equation} 
where $\mathcal{N}_1(n)$ and $\mathcal{N}_2(n)$ are sets of LU and LA nodes, respectively, such that
\begin{eqnarray}
\mathcal{N}_1(n) & =  & \left\{ n':\substack{ \displaystyle 1 \leq n' \leq N \\ \displaystyle n' \neq n}, \|\mathbf{x}_n - \mathbf{x}_{n'} \|_2 \leq d_{max} \right\}  \\
\mathcal{N}_2(n) & =  & \{m: 1 \leq m \leq M, \|\mathbf{x}_n - \mathbf{a}_{m} \|_2 \leq d_{max} \} 
\end{eqnarray}
and 
\begin{eqnarray}
d_{nk} = \begin{cases} \|\mathbf{x}_n - \mathbf{x}_k \|_2 \ \ \ \mbox{if } k \in \mathcal{N}_1(n) \\
\|\mathbf{x}_n - \mathbf{a}_k \|_2 \ \ \ \mbox{if } k \in \mathcal{N}_2(n) \end{cases}
\end{eqnarray}

\subsubsection*{Problem to be solved}
Given the inexact LA positions $\mathbf{\bar{a}}_m \in \mathbf{R}^2$, their error upper bound $\epsilon$, and the estimated distance between the nodes, $\bar{d}_{nk}$, where $n = 1, \ldots, N$ and $k \in \mathcal{N}(n)$, estimate the positions of the LU nodes $\mathbf{x}_n \in \mathbf{R}^2$.

\section{The Optimization Problem}
The estimation of the LU node positions $\mathbf{x}_n$ can be formulated as an error minimization problem given by~\cite{biswas}
\begin{eqnarray}
\mbox{minimize } & & \sum_{n=1}^{N-1} \sum_{\substack{n' \in \mathcal{N}_1(n)  \\ n' > n}}  | \| \mathbf{x}_n - \mathbf{x}_{n'}\|_2^2 - \bar{d}^2_{nn'}| \notag   \\
& & + \sum_{n=1}^{N} \sum_{n' \in \mathcal{N}_2(n)}  | \| \mathbf{x}_n - \mathbf{a}_{n'}\|_2^2 - \bar{d}_{nn'}^2|  \label{optProblem1}
\end{eqnarray}
The optimization problem in (\ref{optProblem1}) is nonconvex. However, using the SDR method as in~\cite{biswas}, the problem can be converted to a convex optimization problem as
\begin{eqnarray}
\mbox{minimize } & &  \xi  \label{optProblem2} \\
\mbox{subject to: } & & Y \succeq X^T X \notag 
\end{eqnarray}
where
\begin{eqnarray}
\xi & = & \sum_{n=1}^{N-1} \Bigl( \sum_{\substack{n' \in \mathcal{N}_1(n)  \\ n' > n}} | A^T_{nn'} D A_{nn'}  - \bar{d}_{nn'}^2|   \notag \\
& & \ \ \ \ \ +  \sum_{n' \in \mathcal{N}_2(n)} | B^T_{nn'} D B_{nn'} - \bar{d}_{nn'}^2|  \Bigr) \label{xiVal}\\
D & = &  \left[ \begin{matrix} Y & X^T \cr X & I_2 \end{matrix} \right]  \\
A_{nn'} & = & [\mathbf{e}_{nn'}^T ~\mathbf{0}_2^T]^T \\
B_{nn'} & = & [\mathbf{e}_{n}^T ~\mathbf{a}_{n'}^T]^T \\
\mathbf{e}_{nn'} & = & \mathbf{e}_{n} - \mathbf{e}_{n'} \\
X & = & [\mathbf{x}_1 \ldots \mathbf{x}_N ]^T
\end{eqnarray}
$\mathbf{e}_n \in \mathbf{R}^N$ is the $n$th unit vector, $\mathbf{0}_2 \in \mathbf{R}^2$ is a zero vector, $\mathbf{I}_2 \in \mathbf{R}^{2\times 2}$ is an identity matrix, and $X \in \mathbf{R}^{2 \times N}$, $Y \in \mathbf{R}^{N \times N}$ are optimization variables. 

The optimization problem in (\ref{optProblem2}), however, does not work well when there are errors in the distance estimates and in the positions of the LA nodes~\cite{biswas}. In the following subsection, we describe an optimization method that yields more accurate node positions and performs well even in the presence of node-position and node-distance errors.

\subsection{The Proposed Method}
The optimization problem in (\ref{optProblem1}) attempts to estimate the positions of the LU nodes by ensuring that the distances between the nodes are as close as possible to the given values. In cases where an LU-node position that satisfies the problem in (\ref{optProblem1}) is not unique, the estimate of the node position using SDR in (\ref{optProblem2}) will have an error. However, this error usually reduces as the number of non-unique positions becomes smaller and becomes zero when the node position is unique.

In our proposed method, we introduce a regularization term $\zeta$ that penalizes an LU node if it is close to another node that has no direct connection with it. The term is defined as
\begin{equation}
\zeta = -\sum_{n=1}^{N-1} \left( \sum_{\substack{n' \notin \mathcal{N}_1(n)  \\ n' > n}}  \| \mathbf{x}_n - \mathbf{x}_{n'}\|_2^2 + \sum_{n' \notin \mathcal{N}_2(n)}  \| \mathbf{x}_n - \mathbf{a}_{n'}\|_2^2 \right)
\label{regTerm}
\end{equation}
Since the above term favors certain configurations over others it therefore helps to reduce the number of non-unique positions. Using SDR, the regularization term in (\ref{regTerm}) can be approximated as a convex formulation given by
\begin{equation}
\hat{\zeta} = -\sum_{n=1}^{N-1} \left( \sum_{\substack{n' \notin \mathcal{N}_1(n)  \\ n' > n}}  A^T_{nn'} D A_{nn'} + \sum_{n' \notin \mathcal{N}_2(n)} B^T_{nn'} D B_{nn'} \right)
\end{equation}
It should be pointed out that the term $\hat{\zeta}$ is quite different from the regularization term in~\cite[eqn.~(16)]{biswas}, which makes no distinction whether or not a node is directly connected to another node and, furthermore, requires an additional SDR evaluation to determine the weighting factor. While the goal of the term in~\cite{biswas} was to prevent the estimated nodes from crowding together when the data is noisy, the term $\hat{\zeta}$ is meant to reduce or eliminate the non-unique positions by giving more weightage to certain configurations, regardless of whether or not the data is noisy. 

If the number of non-unique solutions of the LU positions is small (e.g., less than 3 independent solutions), the use of the regularization term $\hat{\zeta}$ in the optimization has been found to be quite effective in removing most of the ambiguity thereby leading to the correct solution. However, if the network is not well connected, the number of non-unique solutions will not be small and the use of $\hat{\zeta}$ is not as effective. More specifically, in a poorly connected network where the number of non-unique solutions is quite high, the use of $\hat{\zeta}$ can be counter-productive: this is because the regularization term in the optimization will not be able to remove all the ambiguity. Consequently, from the remaining ambiguous positions the regularization term will tend to give more preference to the linear combinations where the nodes that are not directly inter-connected are far apart, leading to a solution with a large error in most cases. 

Therefore, to decide when to employ $\hat{\zeta}$ in the optimization problem, we introduce a measure of the connectivity of the LU nodes, given by
\begin{equation}
\mathcal{C} = \frac{\displaystyle \sum_{n=1}^N  \left( \displaystyle |\mathcal{N}_1(n)|+|\mathcal{N}_2(n)| \right)}{N^2+NM}
\end{equation}
where $0 \leq \mathcal{C} \leq 1$. A higher value of $\mathcal{C}$ implies that the LU nodes are more connected within the network and, therefore, more weightage can be given to $\hat{\zeta}$ in the optimization. Consequently, the modified optimization problem is given by
\begin{eqnarray}
\mbox{minimize } & & \xi  +  \kappa~\hat{\zeta}  \label{optProblem3} \\
\mbox{subject to: } & & Y \succeq X^T X \notag 
\end{eqnarray}
where $\kappa(\mathcal{C})$  is the weighting coefficient that is dependent on the network connectivity $\mathcal{C}$. In general, a well connected network will have $\mathcal{C} > 0.7$ and in such cases setting $\kappa(\mathcal{C})$ to $0.1$ gave good results. On the other hand, a poorly connected network usually has $\kappa(\mathcal{C}) < 0.3$ and in such cases, the use of $\zeta$ is not advantageous, so $\kappa(\mathcal{C})$ is set to 0. For values of $\mathcal{C}$  between 0.3 and 0.7, $\kappa(\mathcal{C})$ can be gradually increased using an interpolation function. Consequently, a function of $\kappa(C)$ that gave good results is given by 
\begin{equation}
g(\mathcal{C}) = \begin{cases} 
0 & \mbox{if } \mathcal{C} \leq \Gamma_l \\
c_l & \mbox{if } \Gamma_l \leq \mathcal{C} \leq \Gamma_a \\
c_l + \displaystyle \frac{(c_h - c_l)(\mathcal{C}-\Gamma_a)}{\Gamma_h - \Gamma_a} & \mbox{if } \Gamma_a < \mathcal{C} \leq \Gamma_h \\
c_h &  \mbox{if } \mathcal{C} > \Gamma_h \end{cases}
\label{gVal}
\end{equation}
where $c_l$, $c_h$, $\Gamma_l$, $\Gamma_a$, and $\Gamma_h$ are set to 0.01, 0.1, 0.3, 0.5, and 0.7, respectively. It should be pointed out that though the weighting function in (\ref{gVal}) gives consistently good results for networks with different values of $M$ and $N$, it is obtained empirically by trial-and-error and is, therefore, not the most optimal weighting function. Our future effort is to investigate weighting functions that are statistically optimal with respect to the network connectivity $\mathcal{C}$. 
\section{Experimental Results}
In this section, we provide comparative experimental results to demonstrate the efficiency of the proposed method. For the comparison, we consider three competing methods that are also formulated as convex optimization problems. The first competing method~\cite{chiu} is an SDR formulation that minimizes the worst-case position errors of the LA sensors, the second method~\cite{biswas} is also an SDR formulation, as given by (\ref{optProblem2}), while the third method~\cite{tseng} is an SOCP formulation. Note that in our experiments we only consider networks where each node is directly or indirectly connected to every other node; in other words, the networks have no isolated nodes or isolated subnetworks. 

\begin{table}
\begin{center}
\caption{Typical values of the parameters used in the experiments}
\label{tab_params}
{\footnotesize{
\begin{tabular}{||l|c||} \hline \hline
Parameters  & Values \\ 
\hline 
No. of LU nodes, $N$  & 15   \\
No. of LA nodes, $M$  & 5    \\
Path loss exponent $\gamma_p$  & 3 \\
Standard deviation of fading gain, $\sigma_{dB}$  & 3.5 \\
UB of LA node position error, $\epsilon$ (m) & 0.01 \\
$d_{max}$ (m)  & 0.5  \\
\hline \hline 
\end{tabular} 
}}
\\ \hspace{-1.4in} \footnotesize UB: upper bound
\end{center}
\end{table}

In our experiments, we consider an area of $1\times 1$ $m^2$ where the LU and LA sensors are randomly deployed. The typical values of the parameters for the experiments are tabulated in Table~\ref{tab_params}. In the experiments, we compare the performance of of the different methods by varying the parameters about their typical values. For each trial, the randomly generated locations of the LU and LA nodes are uniformly distributed in a square $[0, 1]\times[0, 1]$, as in~\cite{chiu, biswas}. The inexact LA node position of the $m$th node for the $i$th trial is obtained as
\begin{equation}
\mathbf{\bar{a}}_m^{(i)} = \mathbf{a}_m^{(i)} + {\bm \delta}_m^{(i)}
\end{equation}
where
\begin{equation}
{\bm \delta}_m^{(i)}  =  \epsilon \ [ r_m^{(i)} \cos\theta_m^{(i)}~~ r_m^{(i)}\sin\theta_m^{(i)}]^T
\end{equation}
$\{r_1^{(i)} \ldots r_M^{(i)} \}$ are generated as i.i.d. Gaussian random variables with variance of 1 and mean of 0, and $\{\theta_1^{(i)} \ldots \theta_M^{(i)}\}$ are generated as i.i.d. uniform random variables between 0 and 1.
%\begin{eqnarray}
%{\bm \delta}_m^{(i)} & = & [r_m^{(i)} \cos\theta_m^{(i)}~~ r_m^{(i)}\sin\theta_m^{(i)}]^T \\
%\left[r_1^{(i)} \ldots r_M^{(i)}\right] & = & \epsilon \times randn(2, M) \\
%\left[\theta_1^{(i)} \ldots \theta_M^{(i)}\right] & = & 2\pi \times rand(2, M) \\
%\end{eqnarray}
%and $randn(m, n)$ is defined as an $m\times n$ matrix of normally distributed random variables with a variance of 1 and mean of 0. 

To measure the accuracy of the estimated positions we use the root mean square of the error given by
\begin{equation}
RMSE = \sqrt{\frac{1}{\mathcal{T}} \sum_{i=1}^{\mathcal{T}} E_i^2} = \sqrt{\frac{1}{\mathcal{T}} \sum_{i=1}^{\mathcal{T}} \sum_{n=1}^N \|\mathbf{\hat{x}}_n^{(i)} - \mathbf{x}_n^{(i)} \|_2^2}
\label{EiDef}
\end{equation} 
where $\mathbf{\hat{x}}_n^{(i)}$ is the estimated position of the $n$th LU node during the $i$th trial and $\mathcal{T}$ is the number of trials; in our experiments, $\mathcal{T}$ is set to 50. However, the presence of outliers can sometime over-influence or distort the result in the RMSE. To provide a more detailed perspective of the estimated error and its distribution during the trials, we also display the {\it boxplot}~\cite{mcgill} of the estimated errors $E_1, \ldots, E_{\mathcal{T}}$. The MATLAB command, boxplot, was used for the purpose; for each box, the central mark corresponds to the median, the edges of the box to the 25th and 75th percentiles, and the "+" mark to an outlier, if present; the width of the box is the inter-quartile range of the data.

To compare the performance of the proposed method, we consider four experiments where we independently vary $M$, $d_{max}$, $\sigma_{dB}$, and $\epsilon$ in each of them. For the parameters that are not changing, the values given in Table~\ref{tab_params} were used.
\subsection{Experiments 1 and 2}
\begin{figure}
\begin{center}
\epsfxsize=8.5cm
\epsfbox{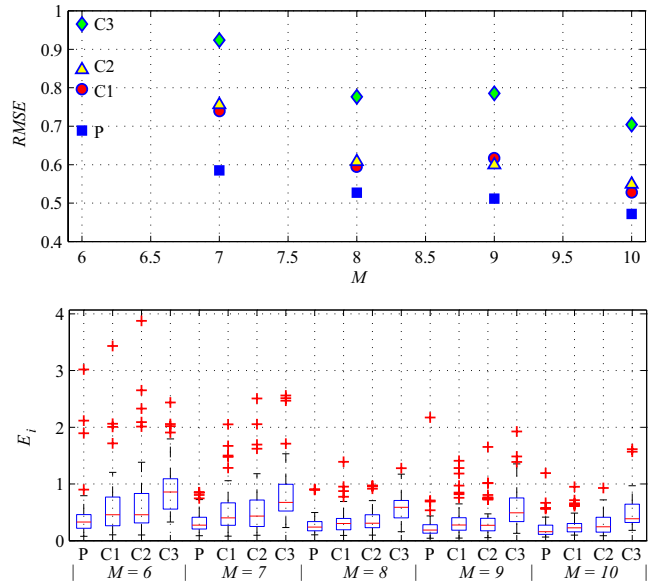}
\caption{Comparisons of the $RMSE$ (top) and the boxplots of $E_i$ (bottom) at different values of $M$ for Experiment 1 (P: proposed method; C1: method in~\cite{chiu}; C2: method in~\cite{biswas}; C3: method in~\cite{tseng}). The error $E_i$ is defined in (\ref{EiDef}).}
\label{fig1}
\end{center}
\end{figure}

\begin{figure}
\begin{center}
\epsfxsize=8.5cm
\epsfbox{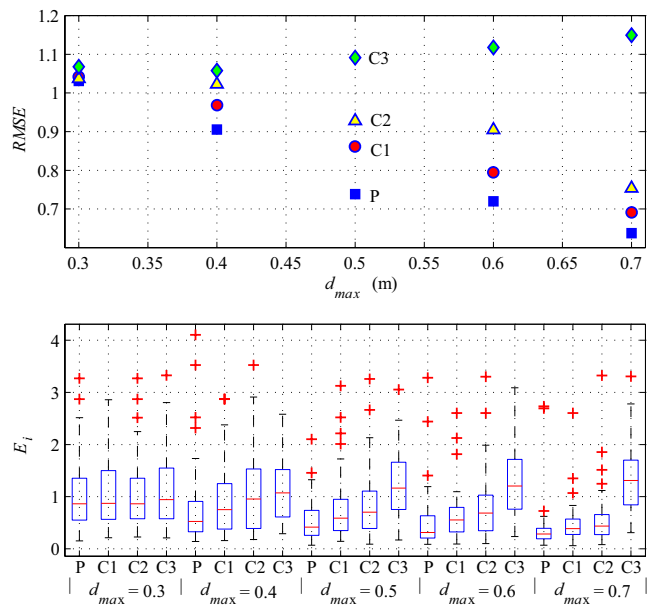}
\caption{Comparisons of the $RMSE$ (top) and the boxplots of $E_i$ (bottom) at different values of $d_{max}$ for Experiment 2 (P: proposed method; C1: method in~\cite{chiu}; C2: method in~\cite{biswas}; C3: method in~\cite{tseng}). The error $E_i$ is defined in (\ref{EiDef}).}
\label{fig2}
\end{center}
\end{figure}

In Experiments 1 and 2, we compared the performance of the proposed method with the competing methods for different values of $M$ and $d_{max}$, respectively. The comparison plots of the $RMSE$ and boxplots of the estimated error, $E_i$, for Experiment 1 and Experiment 2 are shown in Figs.~\ref{fig1} and \ref{fig2}, respectively.
As can be seen from the plots, the sensor positions computed using the proposed method have the smallest $RMSE$ in both the experiments. From the boxplots, we observe that the proposed method has the smallest median values among the three methods. In addition, it also has the smallest inter-quartile range for most of the test cases. It is interesting to note in Fig.~\ref{fig2} that the $RMSE$ and median values of the four methods are close to one another when $d_{max}$ is relatively small at 0.3 m. A possible reason is due the low inter-connectivity between the nodes when $d_{max}$ is small, thereby increasing the existence of higher number of non-unique solutions that satisfy the optimization problem in (\ref{optProblem1}); this, in turn, degrades the accuracy of the estimated solution for all the four methods.
\subsection{Experiments 3 and 4}
\begin{figure}
\begin{center}
\epsfxsize=8.5cm
\epsfbox{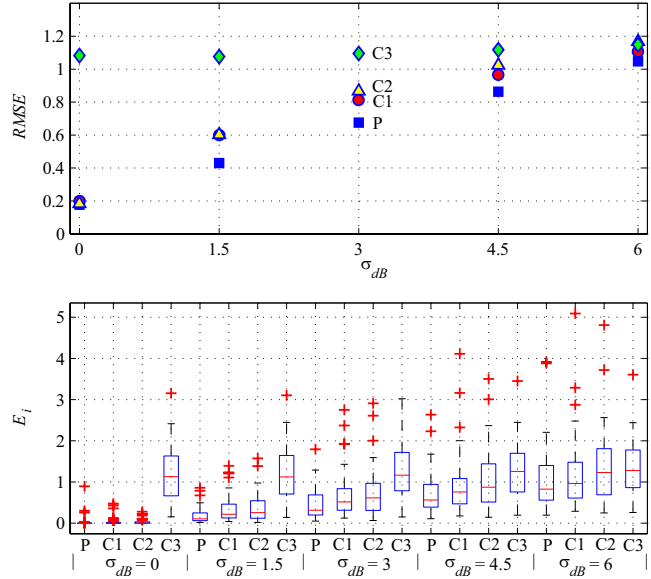}
\caption{Comparisons of the RMSE (top) and boxplots of $E_i$ (bottom) at different values of $\sigma_{dB}$, for Experiment 3 (P: proposed method; C1: method in~\cite{chiu}; C2: method in~\cite{biswas}; C3: method in~\cite{tseng}). The error $E_i$ is defined in (\ref{EiDef}).}
\label{fig3}
\end{center}
\end{figure}

\begin{figure}
\begin{center}
\epsfxsize=8.5cm
\epsfbox{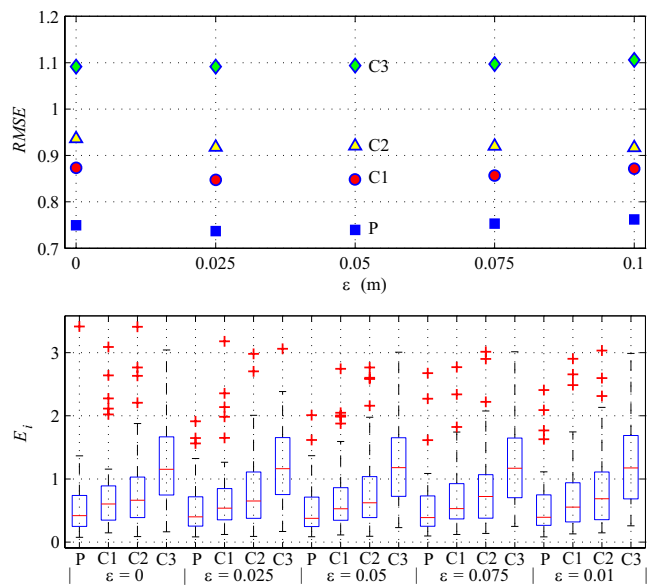}
\caption{Comparisons of the RMSE (top) and boxplots of $E_i$ (bottom) at different values of $\epsilon$ for Experiment 4  (P: proposed method; C1: method in~\cite{chiu}; C2: method in~\cite{biswas}; C3: method in~\cite{tseng}). The error $E_i$ is defined in (\ref{EiDef}).}
\label{fig4}
\end{center}
\end{figure}

In Experiments 3 and 4, we compared the performance of the proposed method with the competing methods for different values of the fading-gain standard deviation, $\sigma_{dB}$, and LA-position error, $\epsilon$, respectively. Note that an increase in the fading gain implies an increase in the measurement error of the distance between the nodes as given by the relation in (\ref{distFading}). The comparison plots of the $RMSE$ and boxplots of the estimated error, $E_i$, for both the experiments are included in Figs.~\ref{fig3} and \ref{fig4}, respectively. 
From the RMSE plot in Fig.~\ref{fig3}, we observe that the proposed method has the smallest RMSE among all the methods. The boxplot in Fig.~\ref{fig3} gives a more informative picture that clearly shows the effectiveness of the proposed method across different values of the fading gain $\sigma_{dB}$. In Fig.~\ref{fig4}, we observe from both the RMSE and boxplots that the proposed method yields more accurate node-position estimates than the competing methods whether the LA node positions are exact or have varying levels of errors. Consequently, from the results of Examples 3 and 4, we can conclude that the proposed method is more robust to errors in the distances between the nodes and the LA node positions than the competing methods.

The above experiments have shown that the proposed method yields LU-node positions with the smallest $RMSE$ and median error compared to those achieved with the competing methods considered. It should be pointed, however, that there exists a small percentage of node configurations where the proposed method or the competing methods yield poor solutions, as indicated by the outliers in the boxplots. In our future work, we plan to study the configurations of the outliers more closely, and to investigate techniques to detect such configurations, including optimization methods to solve them. 
  
\section{Conclusions}
A new method for estimating the relative position of LU nodes from the positions of the LA nodes and the received signal strength (RSS) between the nodes, in a wireless sensor network (WSN), has been proposed. In the method, a regularization term is incorporated in the optimization problem leading to significant improvement in the estimation accuracy even in the presence of position errors of the location-aware nodes and distance errors between the nodes. The method is formulated as a convex optimization problem using the SDR approach. Experimental comparisons with state-of-the-art competing methods showed that the proposed method yields node positions with much smaller $RMSE$ and median error, and also performs better in the presence of node-position and node-distance errors.

%\section*{Acknowledgment}
%The authors are grateful to the Natural Sciences and Engineering Research Council of Canada for supporting this work.



%% References


\bibliographystyle{elsarticle-num}
\bibliography{<your-bib-database>}






% \end{thebibliography}

\end{document}